\documentclass{sig-alternate-2013}

\newcommand{\hup}{\vspace*{-0.1cm}}

\newtheorem{defn}{{\bf{Definition}}}

\usepackage{graphicx}
\usepackage{mathtools}

\usepackage{amsmath,amssymb,amsfonts}

\usepackage{url}
\usepackage{verbatim}
\usepackage{multirow}
\usepackage{caption}
\usepackage{subcaption}

\usepackage[linesnumbered,ruled,vlined]{algorithm2e}
\usepackage{algpseudocode}
\usepackage{mathtools}

\usepackage{ifpdf}
\usepackage{lmodern}
\usepackage{lipsum}
\usepackage[T1]{fontenc}
\usepackage{textcomp}
\usepackage{enumitem}
\usepackage{mdwlist}
\usepackage{pbox}
\usepackage{comment}
\usepackage{ifthen}
\usepackage{color}
\usepackage{colortbl,xcolor}
\usepackage{upgreek}
\usepackage{upgreek}
\usepackage{todonotes}

\usepackage{tikz}
\usetikzlibrary{arrows}

\usepackage[hidelinks]{hyperref}

\definecolor{Gray}{gray}{0.9}
\definecolor{LightCyan}{rgb}{0.88,1,1}

\specialcomment{notes}{\begingroup \color{blue}} { \endgroup}

\newcolumntype{!}{>{\global\let\currentrowstyle\relax}}
\newcolumntype{^}{>{\currentrowstyle}}

\newcommand{\si}{\begin{enumerate}}
\newcommand{\ii}{\item}
\newcommand{\ei}{\end{enumerate}}

\makeatletter
\let\oldfootnote\footnote
\def\footnote{\@ifstar\footnote@star\footnote@nostar}
\def\footnote@star#1{{\let\thefootnote\relax\footnotetext{#1}}}
\def\footnote@nostar{\oldfootnote}
\makeatother

\DeclareGraphicsExtensions{.eps,.png}
\graphicspath{{figure/}{figure/eps/}{figure/png/}}

\newcommand{\superscript}[1]{\ensuremath{^{\textrm{#1}}}}

\makeatletter
\def\@copyrightspace{\relax}
\makeatother

\begin{document}

\title{When-To-Post on Social Networks}

\def\kloutinc{\superscript{*}}
\def\rutgers{\superscript{\dag}}

\numberofauthors{1} 
\author{
    \alignauthor Nemanja Spasojevic, Zhisheng Li, Adithya Rao, Prantik Bhattacharyya\\
    \affaddr{Lithium Technologies | Klout}  \\
    \affaddr{San Francisco, CA}  \\
    \email{\{nemanja, zhisheng.li, adithya, prantik\}@klout.com}
} 

\maketitle

\begin{abstract}
For many users on social networks, one of the goals when broadcasting content is to reach a large audience.  
The probability of receiving reactions to a message differs for each user and depends on various factors, such as location, daily and weekly behavior patterns and the visibility of the message.
While previous work has focused on overall network dynamics and message flow cascades, the problem of recommending personalized posting times has remained an under-explored topic of research.

In this study, we formulate a \textit{when-to-post} problem, where the objective is to find the best times for a user to post on social networks in order to maximize the probability of audience responses.
To understand the complexity of the problem, we examine user behavior in terms of post-to-reaction times, and compare cross-network and cross-city weekly reaction behavior for users in different cities, on both Twitter and Facebook.
We perform this analysis on over a billion posted messages and observed reactions, and propose multiple approaches for generating personalized posting schedules.
We empirically assess these schedules on a sampled user set of $0.5$ million active users and more than $25$ million messages observed over a $56$ day period.
We show that users see a reaction gain of up to $17\%$ on Facebook and $4\%$ on Twitter when the recommended posting times are used.

We open the dataset used in this study, which includes timestamps for over $144$ million posts and over $1.1$ billion reactions.
The personalized schedules derived here are used in a fully deployed production system to recommend posting times for millions of users every day.
\end{abstract}



\section{Introduction}
\label{section:introduction}
Social networks have emerged as major platforms for communication in recent years, with hundreds of millions of interactions created by users every day.
Though the underlying mechanisms may vary, a large number of active interactions may be classified under (a) users posting messages, or (b) users reacting to messages. 
Posted messages may sometimes be intended for a few friends and family members, while other times they may be geared towards larger audiences.
The latter is especially true for users such as brands, marketers and public figures, who leverage social media as platforms for broadcasting messages.

One of the goals while broadcasting messages is to capture the attention of audience members so that they may react to the posted message.
The probability that an audience member reacts to a message may depend on several factors, such as his daily and weekly behavior patterns, his location or timezone, and the volume of other messages competing for his attention.
The problem of broadcasting messages at the right time in order to elicit responses from one's audience is therefore a complex one with many dimensions.

A large body of research in this area has focused on the problem of influence maximization and related topics, where the goal is to target a specific subset of users in order to create information cascades in the network.
However, the dynamics of broadcasting to entire audiences, rather than picking specific individuals to target, has been an under-explored topic of study. 
Further, since each user has a unique audience, any recommendations for posting times need to be personalized to be effective, as we show in this study.
We hence formulate a \textit{when-to-post} problem here, where the objective is to find the best times for a user to post on social networks in order to increase audience responses.  


Apart from introducing the problem, our contributions in this work are three-fold.
First, in order to understand the complexity of the \textit{when-to-post} problem and the factors that affect it, we perform in-depth \textbf{user reaction behavior analysis}, which includes:
\si 
\ii \textit{Post-to-reaction behavior: } We analyze the delays between posting and reaction times across different social networks and user in-degrees.
\ii \textit{Cross-network analysis: }We examine the similarities and differences of audience behavior on Twitter and Facebook. 
\ii \textit{Cross-city analysis: }We compare cycles of daily and weekly user activity in different cities, and present analysis on how location affects posting schedules.
\ei

Second, we formally define the \textit{when-to-post} problem in a probabilistic setting, and propose multiple approaches for \textbf{recommending personalized posting schedules}.
Among these are the \textit{First-Degree} and the \textit{Second-Degree} schedules, and their corresponding weighted counterparts. 
We empirically assess these schedules against two global baselines, on a real-world set of $0.5$ million active users observed over a $56$ day period.
We define a metric called \textit{Reaction Gain} that helps us evaluate the effectiveness of the two approaches, and show that users see an average reaction gain of up to $17\%$ for Facebook and upto $4\%$ for Twitter.

Third, we open a public dataset consisting of anonymized user ids and timestamp data that could help future research in this area. This dataset contains timestamps for \textbf{144 million} posts and \textbf{1.1 billion} reactions from a $120$-day period.

We performed our study and analysis on a full production system deployed on \textit{klout.com}. 
Klout\footnote{Klout platform is a part of Lithium Technologies, Inc.} is a social media platform that aggregates and analyzes data from social networks \cite{nemanja-lasta} such as Twitter, Facebook, Google+ and others.
Our system recommends personalized posting schedules for millions of users to share content on Twitter and Facebook.


\section{Related Work}
\label{section:related}

The subject of user behavior dynamics on social networks has attracted significant research attention \cite{ Lehmann2012DCC, crane2008robust, Bennamane2011VAI}.
Wu et al.~\cite{wu2011says} categorized Twitter users into elite and casual users and analyzed the differences in how they generate and consume information.
In their study, they showed that regardless of the type of content, all content had very short life spans that usually dropped exponentially after a day.
Another study in \cite{asur2011trends} also showed that only a few topics lasted for a long time on social media platforms, while most topics faded away quickly in the order of 20-40 minutes.

Besides the life span of messages, researchers have also analyzed the effects of timezone and location on user activity patterns.
Kwak et al.~\cite{kwak2010twitterwhat} analyzed the timezone characteristics of user audiences on Twitter and reported that
the average timezone difference between a user and her friends varied with the number of friends.
In our study, we further analyze the impact of audience location on the volume of responses towards a message.

There have been several studies on modeling the dynamics of social network events \cite{Leskovec2009MDN, tsytsarau2014dynamics}. For example, the work in \cite{tsytsarau2014dynamics} used different convolution functions to analyze the flow of news events and sentiments through Twitter.
While the approach of these studies has been to analyze the overall temporal characteristics on social media, here we take the further step of analyzing reaction behavior from the point of view of each individual user, thereby enabling personalized recommendations for posting messages.
%


Another line of related research is in the area of information flow and diffusion.
Studies such as \cite{lerman2010information, leskovec2007patterns, cheng2014cascades} have analyzed how factors such as the topological structure of social networks play a role in information cascades.
Yang et al.~\cite{yang2010twitterdiffusion} presented results on analyzing message flow based on Twitter mentions, and found that long-term historical user properties such as the rate of previous mentions were as important as the tweet content.
The authors in \cite{yang2011temporalvariations} studied the importance of hashtag adoption in determining the popularity and spread of tweets.
The study in \cite{guille2012temporaldynamics} proposed a predictive approach to model dynamics of diffusion in social networks based on social, semantic and temporal dimensions.
However, the problem of examining the flow of messages in the entire network differs significantly from the one in our study. 
Here we are instead concerned with the reactions received by a single user in a short time window.

A large body of research has also focused on influence maximization \cite{Kempe2003KDD, Chen2009KDD, Bharathi2007WINE}, which also differs from the \textit{when-to-post} problem.
Influence maximization aims to find a subset of users in a social network, such that targeting them with a message maximizes the propagation or adoption of the message throughout the network.
However, the effects of broadcasting messages to entire audiences, rather than targeting specific individuals, has not been as well studied. 
It is this problem that we propose and analyze here, by examining the temporal aspects of broadcasting to one's audience, in order to get a large volume of responses.






\section{Problem Setting}
\label{section:problem_statement}
In this section, we formulate the \emph{when-to-post} problem and provide details about the system and dataset used.

\subsection{Problem Statement}
\label{sec:problem_statement}
The actions taken on any social networking site may be categorized as passive or active in nature. 
The passive category may include actions such as \emph{views}, while the active category may broadly be classified into two groups -- \emph{post} and \emph{reaction}. 
Typical \emph{post} behavior may include creating and sending messages, sharing photos, or posting news articles on a social network.
Typical \emph{reaction} behavior includes resharing, liking, commenting, endorsing or replying to posts created by other users.
We restrict the scope of this study to the \emph{post} and \emph{reaction} behavior of users.

Sometimes the \emph{post} behavior is used in the context of one-on-one or personal communication, while other times
it may be geared towards a larger audience. Here we focus on the latter case, where one of the motivations behind
posting is to reach a large audience and to capture their attention. In particular, we examine the time-related
aspects of this behavior and frame a \emph{when-to-post} problem as follows:
\begin{description}[leftmargin=0.1em]
\item\textbf{Problem Statement}: \textit{For a user on a social network, find the \textbf{best time to post} a message within a specified time period in order to maximize the probability of receiving audience reactions.}
\end{description}

Note that we only consider first-degree reactions such as replies and retweets on Twitter and comments on Facebook, and not those caused by an audience member resharing the original post.
In other words, we focus mainly on the reactions a post receives by the user's immediate audience, and not on how the post propagates through the network.

\subsection{System Overview}
\label{sec:system_overview}

We collect user posts from Facebook through the oauth-token provided by registered users on Klout.
We also use the oauth-token-based approach to collect the friend graph of users on Facebook and the follower graph for users on Twitter.
Klout partners with GNIP to collect public data generated in the Twitter Mention Stream\footnote{\small \url{https://gnip.com/sources/twitter}}.
For location analysis, we use the city, state and country information provided by registered users on the Klout application.

\begin{figure}[htb]
  \centering
  \fbox{\includegraphics[width=0.85\columnwidth]{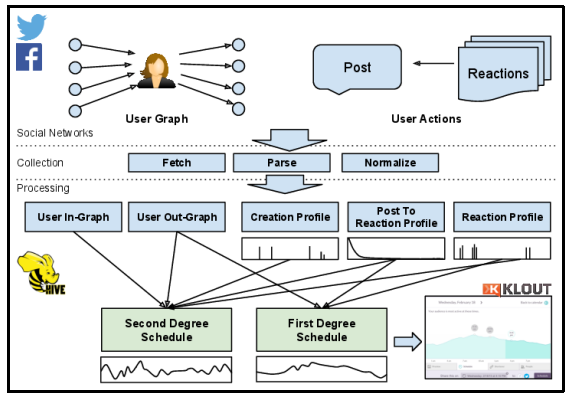}}
  \setlength{\abovecaptionskip}{5pt}
  \caption{System Overview}
  \label{figure:system_overview}
  \vspace{-0.12in}
\end{figure}

The collected data is written out to a Hadoop cluster\footnote{\small \url{http://wiki.apache.org/hadoop/}} that uses HDFS as the file system, HBase as the serving datastore, and Hive\footnote{\small \url{http://hive.apache.org/}} to process, query and manage the large datasets.
We implement independent Java utilities with Hive UDF (User Defined Function) wrappers, with functions to process user locations and timezones, and operators such as discrete convolution to process time-series vectors. 
The combination of Hive Query Language and UDFs allows us to build map-reduce jobs that can scale up to analyze billions of messages posted to social platforms every day.
A pipeline run on a 150-node cluster has a cumulative I/O footprint of 224GB of reads, 78GB of writes, and 9.62 days of CPU usage.
Fig. \ref{figure:system_overview} shows an overview of the system.

\subsection{Dataset}
\label{sec:timestamp_dataset}

The dataset used to run experiments and build models has been opened at \textbf{\url{https://github.com/klout/opendata}}.
The corpus has event timestamps for posts that were created between October 15, 2014 and February 11, 2015 and received at least one reaction.
The dataset was generated from more than \textbf{1 million} users apiece from Facebook and Twitter, with accounts registered on Klout.com.
For Facebook the dataset includes more than \textbf{25 million} post timestamps and \textbf{104 million} reaction timestamps, while for Twitter these numbers are \textbf{119 million} and \textbf{1 billion} respectively.
In order to preserve privacy, timestamps were slightly perturbed and user and post ids were \textbf{anonymized} using custom fingerprint functions.


\section{Behavior Analysis}
\label{section:behavior_analysis}
In this section we perform in-depth user behavior analysis across temporal and local dimensions, such as post-to-reaction delay, user location and the network of activity. 
This analysis provides some interesting observations and valuable insights into the \textit{when-to-post} problem. 

 %
\subsection{Post to Reaction Time Analysis}
\label{subsection:ptr_analysis}

%
%


\begin{figure}[htbp]
  \centering
  \includegraphics[width=0.95\columnwidth]{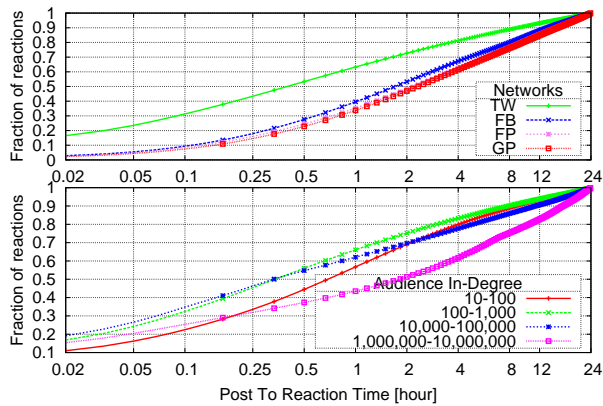}
\vspace{-0.05in}  
  \caption{\small{Cumulative Reactions within first 24 hours}}
  \label{fig:post_to_response}
\vspace{-0.05in}  
\end{figure}

To start with, we note that there is always an inherent delay between when a post was created and when a user reacts to it. 
This delay is crucial to consider when we study the \textit{when-to-post} problem. 

\begin{table}
\begin{center}
  \begin{tabular}{|c||c|c|c|c|} 
  \hline
  \rowcolor{Gray} & \multicolumn{4}{c|}{\textbf{Networks}}  \\ \hline
  \rowcolor{Gray} $\mathcal{T}_{24h}(p)$ & {TW} & {FB} & {FP} & {GP} \\ \hline
    $0.25$ &  00:03 &  00:25 &  00:31 &  00:35 \\ \hline
    $0.50$ &  00:24 &  01:42 &  02:12 &  02:19 \\ \hline
    $0.75$ &  02:24 &  05:65 &  07:26 &  07:36 \\ \hline
    $0.90$ &  08:53 &  13:14 &  14:57 &  15:16 \\ \hline
  \rowcolor{Gray} & \multicolumn{4}{c|}{\textbf{Audience In-Degree (Twitter)}} \\ \hline
  \rowcolor{Gray} $\mathcal{T}_{24h}(p)$ & {10-100} & {100-1K} & {10K-100K} & {1M-10M} \\ \hline
    $0.25$ &  00:08 &  00:03 &  00:03 &  00:06 \\ \hline
    $0.50$ &  00:41 &  00:20 &  00:20 &  01:48 \\ \hline
    $0.75$ &  02:53 &  01:58 &  03:11 &  07:52 \\ \hline
    $0.90$ &  08:49 &  07:50 &  11:22 &  16:26 \\ \hline
  \end{tabular}
  \\
\end{center}
\vspace{-0.2in}
\caption{$\mathcal{T}_{24h}(p)$, Post-to-Reaction Times [hh:mm]}
\label{table:post_to_reaction_time}
\vspace{-0.2in}
\end{table}

Specifically, we are concerned with the post-to-reaction delay within a short time window, and we choose this window to be 24 hours.
This is also in accordance with previous studies such as \cite{wu2011says} that have shown that messages on social media are short-lived with exponential dropoff after a day.
In the limiting case when there is no dropoff and the delay is infinite all posts have the same probability of getting responses.
Thus it is because of this dropoff within a finite duration that the \textit{when-to-post} problem becomes important.
Further, since most reactions occur in narrow time windows for both networks, the goal should be to recommend posting times in narrow time buckets.
To examine the speed of reactions, we define a metric $\mathcal{T}_d(p)$ as follows:

\begin{defn}\label{def:tau}
Let $R$ be the total number of reactions received by all posts within a time period $d$ since posting time. Then $\mathcal{T}_d(p)$ is defined as the amount of time that passes between posting time and the time when the cumulative reaction count is equal to a fraction $p$ of $R$. 
\hfill$\Box$
\end{defn}
 
Along with the reaction counts, we use this metric $\mathcal{T}_d(p)$ to further analyze post-to-reaction behavior across different dimensions of the problem. 
Fig. \ref{fig:post_to_response} plots the fraction of cumulative reaction counts occurring within 24 hours of posting and Table \ref{table:post_to_reaction_time} shows the $\mathcal{T}_{24h}(p)$ values respectively.

Further, we would also like to understand the probability distribution of a reaction occurring within a given time window since the time of post creation. 
In order to do this, we define a \textit{Post-to-Reaction Filter} function as follows:
\begin{defn}\label{def:post_to_reaction}
{\bf{Post To Reaction Filter}} For a time interval $d$, the post-to-reaction filter function $PTR(d)$ is defined as a discrete probability distribution over the event that a reaction occurs within time $d$ of creating a post.
\hfill$\Box$
\end{defn}

We estimate the post-to-reaction filter function $PTR(d)$ by aggregating reaction times across all observed messages and reactions in a network. 
This filter function will be used in Sec. \ref{section:derivation} when we derive personalized user schedules.

\subsubsection{Reaction Times By Network}

Posting and reaction behavior varies on social networks because of many factors, such as manner of posting, presentation of posts to users and the set of possible reactions that a user can perform.
We compare post-to-reaction times across three major social networks -- Twitter (TW), Facebook (FB) and Google+ (GP).
We also treat Facebook Fan Pages (FP) as a separate network, since the dynamics of posting and reacting on these pages diverge significantly from personal Facebook pages.
The top halves of Fig. \ref{fig:post_to_response} and Table \ref{table:post_to_reaction_time} show the reaction times for different networks.

We observe that Twitter exhibits a much higher speed of reactions compared to the Facebook.
On Twitter, 25\% of the reactions take place in the first 3 minutes, 50\% within the first half hour, and 90\% within the first 9 hours.
Other networks exhibit slightly slower speeds compared to Twitter, with 50\% of reactions on Facebook, Facebook Pages and Google+ taking place within the first 2 hours of posting.
Interestingly, we see that the Facebook Pages network shows more similar reaction times to Google+ rather than Facebook, indicating that similar responses can be elicited from users belonging to completely disjoint user sets, if the underlying dynamics of interactions are similar.

In the rest of this paper, we mainly focus on Twitter and Facebook, which show significant variations in post-to-reaction delays.
The distribution of post-to-reaction delay for Twitter is narrower and falls off more quickly compared to Facebook.
The $\mathcal{T}_{24h}(p)$ values in Table \ref{table:post_to_reaction_time} suggest that a 15 minute bucket can capture the necessary granularity of reactions, which we choose as the length of our time buckets. 

These variations also highlight that social networks operate on different timescales, and the post-to-reaction filter function needs to be computed separately for each network during comparison. 
Next, we consider the dependence of reaction behavior on the in-degree of users posting messages. 

\subsubsection{Reaction Times By User In-Degrees}

Next, we explore the hypothesis that network sizes of users may be a factor that affects reaction times.
To do so, we analyze how an audience member's in-degree affects his reaction behavior.
Fig. \ref{fig:post_to_response} (bottom) plots the fractions of 24 hour reaction counts against the time elapsed, for different sets of in-degrees of audience members on Twitter. Table \ref{table:post_to_reaction_time} (bottom) shows the reaction times at various $\mathcal{T}_{24h}(p)$ values. 

We find that a large section of audience members with in-degrees between 100 to 100k exhibit similar behavior.
More than 60\% of the reactions from such users are created in the first 1 hour.
Users with low in-degrees between 10-100 have slower response times, perhaps they may not be very active users.
The users with in-degrees of greater than 1M have the slowest reaction times among all users. 
This may be attributed to such users being celebrities and brands who may not react to messages as quickly as other users do, because of the large volume of messages they see.

Thus, a large portion of audience members show similar reaction behavior, though they may have differing in-degrees.
We can therefore infer that the \textit{when-to-post} problem does not have a large dependency on the network sizes of audience members, unless these sizes are very small or very large. 
This permits us to use a common post-to-reaction filter function for all users in a given network.



\subsection{Network and Location Analysis}

User post and reaction behaviors are multi-dimensional and are highly
dependent on the location, network and timezone of the user.
In this section, we analyze normalized aggregated user audience reaction behaviors $\mathcal{S}(u)$, for user cohorts within and across various cities as well as across Facebook and Twitter within a given city.
For behavior analysis we use correlation and cosine similarity metrics.
Correlation and cosine similarity between finite time series $\mathcal{S}(u_1)$ and $\mathcal{S}(u_2)$ are defined in Equations \ref{eq:correlation} and \ref{eq:similarity} respectively.

Cosine similarity reveals the overlap between time series, while correlation reveals closeness in time dependent patterns between them.
We observe metric distributions for $10$ to $50$ million user pairs, depending on the cohorts compared, where $u_1$ is selected from the first cohort and $u_2$ from the second.
In addition to the metrics above, we compare the raw time series to gain further insights into reaction behaviors in Figs. \ref{fig:post_schedule_analysis_fb} and \ref{fig:post_schedule_analysis_tw}.
\setlength{\abovedisplayskip}{2pt}
\setlength{\belowdisplayskip}{2pt}
\hup%
\begin{equation}\label{eq:correlation}
{\tiny{
corr(\mathcal{S}(u_1),\mathcal{S}(u_2)) =
 \frac{\sum\limits_{i=1}^{N}(s_{u_1,i} - \bar{s}_{u_1}) (s_{u_2,i} - \bar{s}_{u_2})}
 {\sqrt{\sum\limits_{i=1}^{N}(s_{u_1,i} - \bar{s}_{u_1})^2 \sum\limits_{i=1}^{N}(s_{u_2,i} - \bar{s}_{u_2})^2}}}
}
\end{equation}

\setlength{\abovedisplayskip}{2pt}
\setlength{\belowdisplayskip}{2pt}
\hup%
\begin{equation}\label{eq:similarity}
{\tiny{
sim(\mathcal{S}(u_1),\mathcal{S}(u_2)) = \frac{\mathcal{S}(u_1) \cdot \mathcal{S}(u_2)}{||\mathcal{S}(u_1)||  \: ||\mathcal{S}(u_2)||}
}}
\end{equation}

\subsubsection{Network Level Analysis}

\begin{figure}[htbp]
  \centering
  \includegraphics[width=0.95\columnwidth]{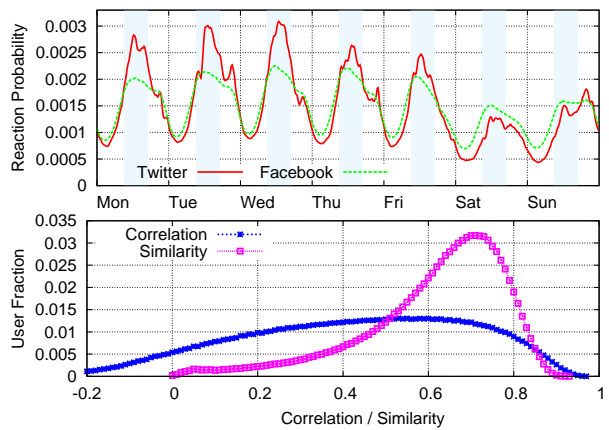}
  \caption{\small{\textit{Top:} Per-Network Globally Aggregated User Audience Reaction Behaviors. \textit{Bottom:} Distribution of Cross-Network Cosine Similarity and Correlation Calculated Per-User. \textit{Both:} All data plotted for users in New York City.}}
  \label{fig:network_analysis}
\vspace{-0.2in}
\end{figure}

In this section, we analyze the user reaction profiles across Twitter and Facebook for users in New York City (NYC).
Fig. \ref{fig:network_analysis} top shows expected audience reactions, aggregated across all users in NYC. 

We observe that the daily seasonality is more pronounced for Twitter than Facebook, with taller peaks and deeper troughs.
Twitter usage seems to peak during working hours and drops quickly thereafter. 
Both networks also exhibit secondary peaks at around 7-8pm daily.
The amplitude of expected reactions on Twitter is around twice that of Facebook's, meaning posting on Twitter at the right times can lead to comparatively larger gains.
Also, compared to Twitter, Facebook usage is more consistent throughout the day.

With respect to weekly trends, we find that Twitter activity falls to almost half of its weekday amplitude on Saturday and Sunday, whereas Facebook activity seems to be less affected by weekends.
It is interesting to note that Facebook is most consistently used throughout the day on Sundays.

We compare aggregated user audience reaction behaviors $\mathcal{S}_{FB}(u_1)$ and $\mathcal{S}_{TW}(u_1)$ for Facebook and Twitter respectively using Eq.~\ref{eq:correlation} and \ref{eq:similarity} in Fig. \ref{fig:network_analysis} bottom.
We observe that correlation is positive, and relatively uniform in the $0.3-0.8$ range, which means that daily audience patterns across Twitter and Facebook are only moderately correlated.
Both the similarity and correlation curves suggests that although audience reactions exhibit some similarity and correlation across networks for a given user, there are still significant differences.
This again reinforces the need for any recommended schedules to be personalized per network.

\subsubsection{Cross-City Analysis}
\label{section:location_analysis}

\begin{figure*}
\centering
\begin{subfigure}{\textwidth}
  \centering
  \includegraphics[width=0.95\textwidth]{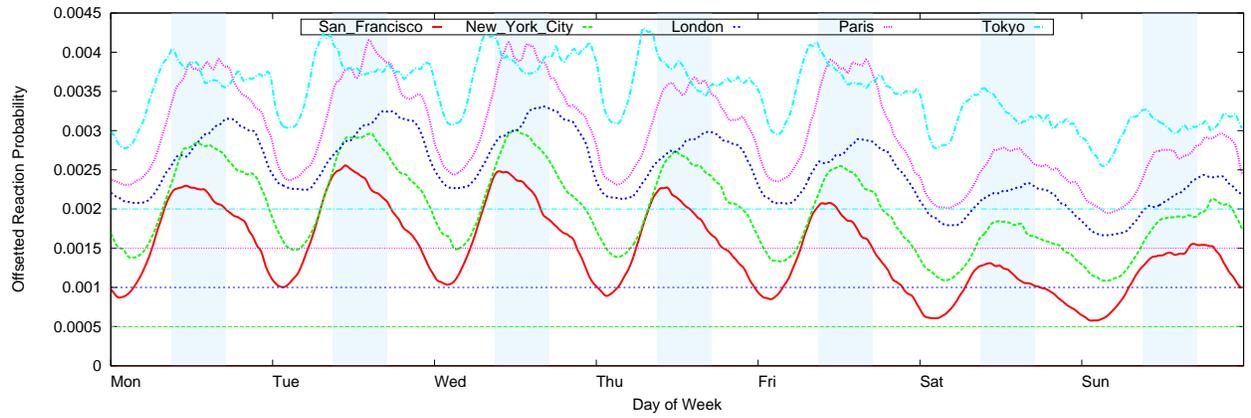}
  \caption{Time Series Capturing City-Level Reaction Behaviors}
  \label{fig:schedule_by_city_fb}
\end{subfigure}

\begin{subfigure}{.24\textwidth}
  \centering
  \includegraphics[height=1.6in]{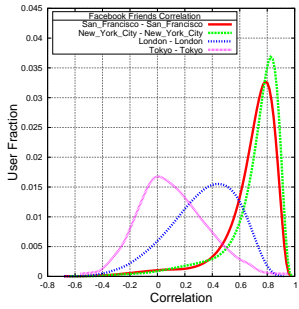}
  \caption{Same-City Correlation}
  \label{fig:same_city_corrleation_fb}
\end{subfigure}
\begin{subfigure}{.24\textwidth}
  \centering
  \includegraphics[height=1.6in]{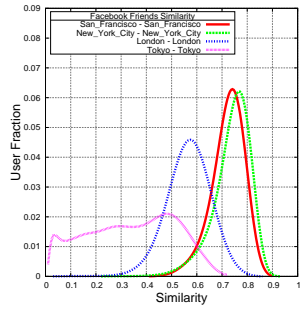}
  \caption{Same-City Similarity}
  \label{fig:same_city_similarity_fb}
\end{subfigure}
\begin{subfigure}{.24\textwidth}
  \centering
  \includegraphics[height=1.6in]{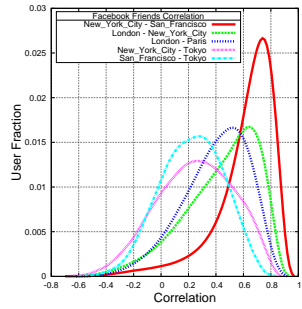}
  \caption{Cross-City Correlation}
  \label{fig:cross_city_corrleation_fb}
\end{subfigure}
\begin{subfigure}{.24\textwidth}
  \centering
  \includegraphics[height=1.6in]{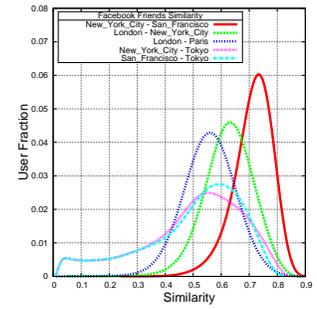}
  \caption{Cross-City Correlation}
  \label{fig:cross_city_similarity_fb}
\end{subfigure}
\caption{\textbf{Facebook} - City-Level Reaction Behavior}
\label{fig:post_schedule_analysis_fb}
\vspace{-0.12in}
\end{figure*}

\begin{figure*}
\centering
\begin{subfigure}{\textwidth}
  \centering
  \includegraphics[width=0.95\textwidth]{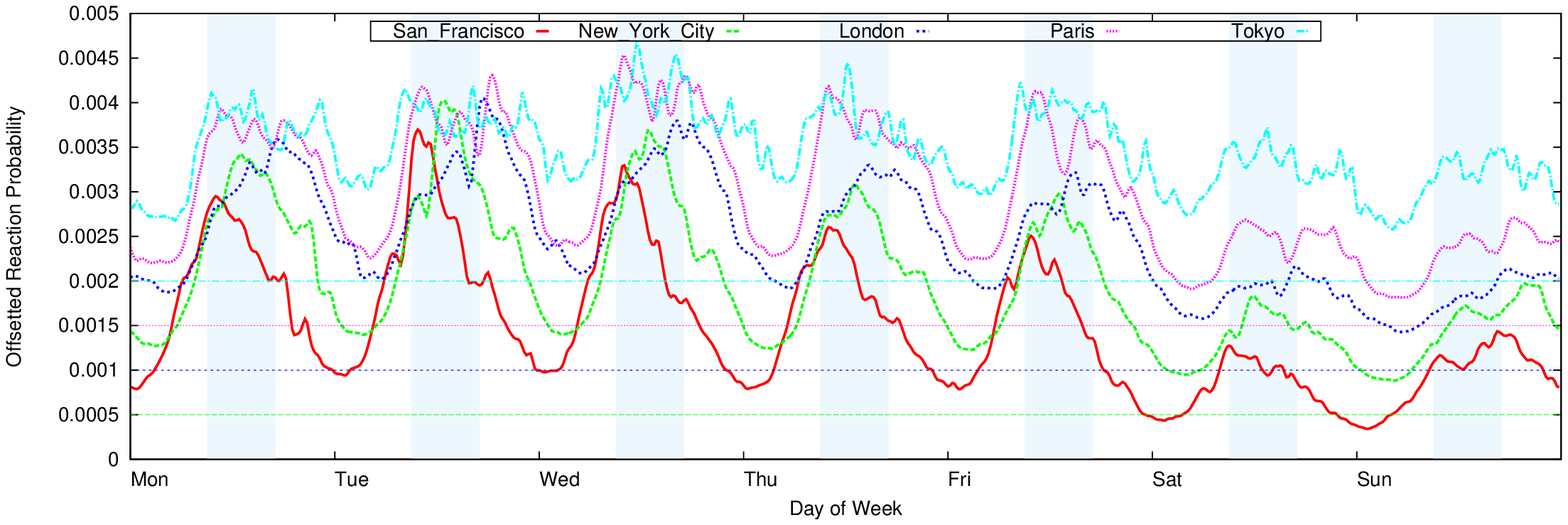}
  \caption{Time Series Capturing City-Level Reaction Behaviors}
  \label{fig:schedule_by_city_tw}
\end{subfigure}

\begin{subfigure}{.24\textwidth}
  \centering
  \includegraphics[height=1.6in]{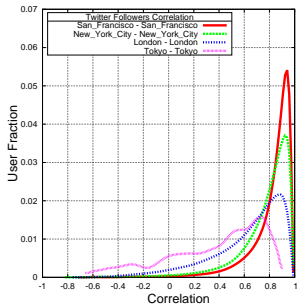}
  \caption{Same-City Correlation}
  \label{fig:same_city_corrleation_tw}
\end{subfigure}
\begin{subfigure}{.24\textwidth}
  \centering
  \includegraphics[height=1.6in]{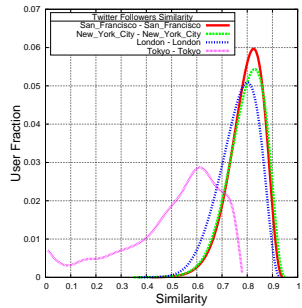}
  \caption{Same-City Similarity}
  \label{fig:same_city_similarity_tw}
\end{subfigure}
\begin{subfigure}{.24\textwidth}
  \centering
  \includegraphics[height=1.6in]{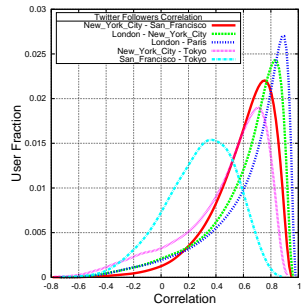}
  \caption{Cross-City Correlation}
  \label{fig:cross_city_corrleation_tw}
\end{subfigure}
\begin{subfigure}{.24\textwidth}
  \centering
  \includegraphics[height=1.6in]{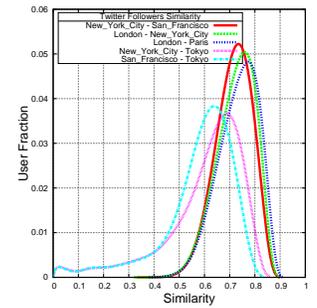}
  \caption{Cross-City Similarity}
  \label{fig:cross_city_similarity_tw}
\end{subfigure}
\caption{\textbf{Twitter} - City-Level Reaction Behavior}
\label{fig:post_schedule_analysis_tw}
\vspace{-0.12in}
\end{figure*}

In this section we analyze differences in behavior for multiple cities across Facebook and Twitter.
Figs \ref{fig:schedule_by_city_fb} and \ref{fig:schedule_by_city_tw} show reaction behaviors,
shifted to the local timezone of the city, for Facebook and Twitter respectively.

Observing the Facebook reactions in Fig. \ref{fig:schedule_by_city_fb}, we notice that the US cities of San Francisco
and New York exhibit similar shapes, where reactions peak at the beginning of work hours.
For Paris, the reactions peak in the second half of working hours, while for London most reactions are expected
towards the end of working hours.
Finally, the pattern for Tokyo is quite different from the rest with two peaks, both occurring off working hours. 

The Twitter reactions in Fig. \ref{fig:schedule_by_city_tw} have similar patterns as Facebook.
The notable difference is that Twitter reactions for US cities have more pronounced daily peaks, while for
London, Paris and Tokyo the behavior seems more consistent throughout the day.
All the curves show significant drops on weekends, and Saturday has noticeably lower activity than Sunday. 
We also observe that New York schedules lag slightly as compared to San Francisco,
which may be explained due to lifestyle differences in the two cities.

In addition to the visual analysis, we also analyze similarity and correlations for reaction behaviors between cities, calculated according to Eq. \ref{eq:correlation} and \ref{eq:similarity}.
The time series compared in this case are the reactions aggregated across users in two cities,
denoted by $\mathcal{S}(\mathcal{C}_1)$ and $\mathcal{S}(\mathcal{C}_2)$.
Figures \ref{fig:same_city_corrleation_fb}-\ref{fig:cross_city_similarity_fb},
\ref{fig:same_city_corrleation_tw}-\ref{fig:cross_city_similarity_tw} show these distributions
for Facebook and Twitter within the same city and across different cities.

Interestingly in US cities (New York and San Francisco) cross-city correlation and similarity
for both Facebook and Twitter are not very different from their within city metrics.
Globally Twitter reaction behavior compared to Facebook seems to be more correlated and similar.
On Facebook, behavior correlation and similarity within city are lowest for London and Tokyo, and have high deviation.
This indicates that users within these cities exhibit more diverse behavior patterns compared to US cities.
Therefore a city level model built for London may not apply to all users within the city.

\section{Personalized Schedules}
\label{section:derivation}
The analysis in the previous section highlights the importance of having personalized posting schedules. Here we present multiple approaches to derive such schedules.

\subsection{Notation and Definitions}
\label{sec:problem_def}

To start with, we simplify the computation by bucketizing time within a period $P$ into discrete time intervals $t_i$.
Based on the analysis in Sec. \ref{section:behavior_analysis}, we use 15 minute time intervals within a period of one week for a total of $4\times24\times7 = 672$ buckets, though the methods described here are applicable to any time interval and period. 
Because the number of reactions in one bucket in each period is usually small for most users, we aggregate the actions from multiple periods into the same bucket. 
For example, all the actions taken by a user between 00:00 to 00:15 on Mondays, in a $90$ day time window, will be grouped into the first bucket $t_1$.

We also define the following sets associated with a user:
\begin{defn}\label{def:out_users}
For a user $u$, the set $U_{out}(u)$ is defined as the set of all users who are connected to $u$, and can potentially
\textbf{react to} the posts created by $u$.
\end{defn}

\begin{defn}\label{def:in_users}
For a user $u$, the set $U_{in}(u)$ is defined as the set of all users to whom $u$ is connected, and whose posts can
be potentially be \textbf{reacted upon} by $u$.
\end{defn}
Note that though we treat the above sets as separate entities in order to differentiate between the post and reaction
behavior, we do not assume that they are disjoint sets.
\footnote{For some bi-directional relationships such as Facebook Friends, $U_{out}(u)$ and $U_{in}(u)$ are equivalent.}

Let $N$ be the number of time buckets within the time period $P$ under consideration. 
To represent the actions associated with a user with respect to time, we create time-based action profiles
for each user computed from a user's actions in the period $P$, and aggregated into the buckets $t_k$.
These profiles can thus be represented as vectors of length $N$.

We define four primary action profiles for each user:
\begin{itemize}
\itemsep0em
   \item First, for each user $u$, we define a \textbf{Created Posts} profile $\mathcal{C}(u)$ that represents the posts created by the user in each time bucket.
   \item Inversely we can also define a \textbf{Visible Posts} profile $\mathcal{V}(u)$, which represents the potentially reactionable posts from $U_{in}(u)$ that are visible to the user.
   \item Based on the posts that a user sees, he may respond to them in some manner. We can represent these responses as a \textbf{Self Reaction} profile $\mathcal{R}(u)$ for the user.
   \item Finally, we define an \textbf{Estimated Audience Reaction} profile $\mathcal{Q}(u)$ that estimates the number of reactions received by the user from his audience $U_{out}(u)$ in each time bucket. 
\end{itemize}






As noted in previous works such as \cite{asur2011trends} and \cite{tsytsarau2014dynamics}, and as analyzed in Sec. \ref{section:behavior_analysis}, there is usually a time difference between when a post is created by a user in $U_{in}(u)$, and when the user $u$ may react to it.
Thus a specific post may be visible in the time bucket $t_k$ in $\mathcal{V}(u)$, but may only be reacted upon in a later time bucket $t_{k^\prime}$ in $\mathcal{R}(u)$.
The post-to-reaction filter function defined in the previous section represents this lag in terms of a time interval $d$, discretized into time buckets of size $t_k$. 
We can therefore compute a \textbf{Delayed Reaction Profile} for a user by performing a discrete convolution operation of the original reaction profile with the post-to-reaction filter function.
\setlength{\abovedisplayskip}{2pt}
\setlength{\belowdisplayskip}{2pt}
\begin{equation}
\mathcal{R}_d(u) = \mathcal{R}(u) \ast PTR(d)
\end{equation}
where $\ast$ is the discrete convolution operator.\footnote{For two functions $f$, $g$ defined on the set of integers $\mathbb{Z}$, the discrete convolution of $f$, $g$ is given by: \\
$(f \ast g)[n] = \sum\limits_{m=-\infty}^{\infty} f[m] \cdot g[n-m]$}

Each element $r_{d,k}(u)$ in the delayed reaction profile represents the number of reactions that the user $u$ would generate in the time interval $d$ following the bucket $t_k$.
Thus for a post created by a user in the current time bucket, using $\mathcal{R}_d(u)$ for his audience members provides a better estimate of anticipated future reactions.




These estimates for $q_i(u)$ could be computed in multiple ways, as described in the following section. Once $\mathcal{Q}(u)$ is known, we can determine a probability mass function which represents a post schedule for the user.
These probabilities $s_i(u)$ can be computed as:
\setlength{\abovedisplayskip}{2pt}
\setlength{\belowdisplayskip}{2pt}
\begin{equation}\label{equation:d_to_s}
s_i(u) = q_i(u) \big/ \sum\limits_{j=1}^{N} q_j(u) \\
\end{equation}

Finally, the vector consisting of these probabilities determine the \textbf{Post Schedule} for the user.
Once we have $\mathcal{S}(u)$, we simply pick the buckets with the highest values of $s_i(u)$, which are the desired best times to post.
Next, we describe multiple approaches to compute $\mathcal{S}(u)$ using the above notation and definitions, which are summarized in Table \ref{action-profiles}.
\begin{table*}[htdp]
\caption{Notation for Action Profiles}
\vspace{-0.2in}
\begin{center}
\begin{tabular}{|p{4.5cm}|p{1.5cm}|p{1.5cm}|p{8.5cm}|}
\hline
\rowcolor{Gray} \textbf{User Action Profile} & \textbf{Vector Notation} & \textbf{Element Notation} & \textbf{Element Description for user $u$ in time bucket $t_k$}\\
\hline
Created Posts & $\mathcal{C}(u)$ & $c_k(u)$ & aggregated number of posts created by user \\
\hline
Visible Posts & $\mathcal{V}(u)$ & $v_k(u)$ & aggregated number of posts visible to user  \\
\hline
Self Reactions & $\mathcal{R}(u)$ & $r_k(u)$ & aggregated number of reactions generated by user \\
\hline
Delayed Self Reactions & $\mathcal{R}_d(u)$ & $r_{d,k}(u)$ & aggregated number of reactions generated by user in the time interval $d$ following $t_k$  \\
\hline
Estimated Audience Reactions & $\mathcal{Q}(u)$ & $q_k(u)$ & estimated number of reactions received by user  \\
\hline
Post Schedule & $\mathcal{S}(u)$ & $s_k(u)$ & probability of receiving a reaction on a post created by user \\
\hline
\end{tabular}
\vspace{-0.2in}
\end{center}
\label{action-profiles}
\end{table*}%

\subsection{Recommended Schedule Derivation}
\label{sec:derivation}

\begin{figure}[htbp]
\begin{tikzpicture}
\tikzset{vertex/.style = {shape=circle,draw,minimum size=0.5em}}
\tikzset{edge/.style = {->,> = latex'}}
\tikzset{main node/.style={circle,draw,,fill=gray!20,minimum size=1.0em}}

  \node[main node] (a) at  (2,0) {\textbf{$a_0$}};
  \node[main node] (b) at  (5,0) {\textbf{$b_0$}};
  \node[vertex] (b1) at  (0,0.5) {$b_1$};
  \node[vertex] (bm) at  (0,-0.5) {$b_m$};
  \node[vertex] (a1) at  (7,0.5) {$a_1$};
  \node[vertex] (an) at  (7,-0.5) {$a_n$};

  \draw[edge] (a) to node[above] {Post} (b);
  \draw[edge] (a) to node[above right] {$U_{out}(a_0)$} (b1);
  \draw[edge] (a) to (bm);
  \draw[edge] (a1) to node[above left] {$U_{in}(b_0)$} (b);
  \draw[edge] (an) to (b);
  \draw[->, red, ultra thick] (b) to[bend left] node[below] {Reaction}  (a);

  \path (a1) -- (an) node [midway, sloped] {$\dots$};
  \path (b1) -- (bm) node [midway, sloped] {$\dots$};
\end{tikzpicture}
\centering
\caption{Simplified representation of a user's social graph}
\label{fig:social_graph}
\end{figure}
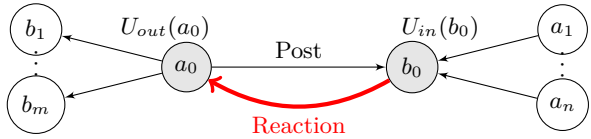

To illustrate the \textit{when-to-post} problem with a concrete example, consider a simplified social network graph,
as represented in Fig. \ref{fig:social_graph}.
For the user $a_0$, her audience is made up of other users $b_i$, so we have: $U_{out}(a_0) = \{b_0, b_1, b_2, ..., b_m\}$.
 
When $a_0$ creates a post, it may be potentially seen by all the members $b_i$ of her audience. 
Let us focus on a particular audience member $b_0$.
This audience member $b_0$ also belongs to the audience sets of other users $a_i$, and may see posts that are created by each of them.
We can represent this relationship between the users as: $U_{in}(b_0) = \{a_0, a_1, a_2, ..., a_n\}$.

We would like to derive the post schedule $\mathcal{S}(a_0)$ for the user $a_0$. 
In order to do so, we want to answer the following question: \textit{For the user $a_0$, what is the expected number of reactions received from $U_{out}(a_0)$ for a post created in the time bucket $t_k$?}
We describe two approaches below to answer this question and compute the recommended schedule.

\subsubsection{First-Degree Reaction Schedule}
\label{sec:first_degree_derivation}
In this approach, we consider the reactions of $a_0$'s audience $U_{out}(a_0)$, ignoring the second-degree effects of the other posting users $a_i$. 
With respect to Fig. \ref{fig:social_graph}, we consider only the left part of the diagram that represents $a_0$ and $U_{out}(a_0)$ (including $b_0$), and ignore all other $a_i$.

Since we know the reaction profiles $R(b_j)$ for the members of $a_0$'s first-degree graph, we can accumulate these reaction counts per time bucket to get the combined audience reaction profile.
However, since this does not take into account the post-to-reaction delay, a better approach is to aggregate the delayed reaction profiles $R_d(b_j)$ for all $b_j$ in $U_{out}(a_0)$. 

This sum of delayed reactions per bucket gives us the estimated audience reaction profile $Q(a_0)$ for the user, where the elements of the vector are given by:
\setlength{\abovedisplayskip}{2pt}
\setlength{\belowdisplayskip}{2pt}
\begin{equation}\label{equation:first_deg}
q_k(a_0) = \sum\limits_{j=0}^{m} r_{d,k}(b_j)
\end{equation}

Thus in this case, the probability of receiving a reaction in any given time bucket $s_k(a_0)$ can then be computed from $Q(a_0)$ as per Eq. \ref{equation:d_to_s}.
These probabilities determine the \textit{First-Degree Reaction} posting schedule $\mathcal{S}_1(a_0)$.

Note that $\mathcal{S}_1(a_0)$ does not take into account the behavior of an audience member $b_j$ with respect to posts from other users $a_i$. 
In other words, this approach only takes into account the first-degree dependency for the user $a_0$.
We therefore describe another approach that takes into account the second-degree dependency as well.

\subsubsection{Second-Degree Reaction Schedule}
\label{sec:second_degree_derivation}
In Fig. \ref{fig:social_graph}, the actions of the users $a_i$ represent the second-degree effects for user $a_0$, since they affect how $a_0$'s first-degree connection $b_0$ reacts to messages.
To consider these second-degree effects, we define a \textit{Second-Degree Reaction} schedule $\mathcal{S}_2(a_0)$, which can be derived by answering the following questions first, before the original one above.
\begin{itemize}
\itemsep0em
   \item When do the users $a_i$ create posts? 
   \item When does a specific audience member $b_0$ react to the posts created by $a_i$?
   \item What is the probability that $b_0$ reacts to a post in a certain time bucket $t_k$?
\end{itemize}

The answer to the first question is given by the post creation profiles $\mathcal{C}(a_i)$ for each user $a_i$, computed by aggregating the past history of post creation events for the user into time buckets.
To answer the second question, we first compute the reaction profile $R(b_0)$. 
Again, this profile is computed by aggregating the past history of reaction events for $b_0$, which tells us how often he reacts in any given time bucket.
The answer to when $b_0$ reacts with respect to posting times is then given by the delayed reaction profile $R_d(b_0)$, which takes into account the post-to-reaction delay.

For the third question, let $p(b_0, t_k)$ be the probability that user $b_0$ reacts to a post in time bucket $t_k$. 
This event can be modeled as a Bernoulli random variable $X_{b_0,k}$, with the probability of the reaction given
by $p(b_0, t_k)$, thus: 
\setlength{\abovedisplayskip}{5pt}
\setlength{\belowdisplayskip}{5pt}
\begin{equation}\label{equation:expected_x}
E(X_{b_0,k}) = p(b_0, t_k)
\end{equation}
From the point of view of $b_0$, the probability that he reacts to some post in the time bucket $t_k$ depends on the
number of posts that he sees, and his usual reaction behavior in $t_k$\footnote{Since we are concerned only with the time aspects here, we assume that
the posts seen by the user are equally likely to be reacted upon in all other aspects.}.

To estimate the number of posts that are potentially visible to the user $b_0$ in each time bucket, we aggregate the post creation profiles for all $a_i$. 
The number of posts that are actually visible to the user may be modeled as a linear function of the total created posts. 
Thus for a given time bucket $t_k$, the number of posts visible to $b_0$ is given by:
\setlength{\abovedisplayskip}{2pt}
\setlength{\belowdisplayskip}{2pt}
\begin{equation}
v_k(b_0) = \alpha \cdot \sum\limits_{i=0}^{n} c_k^\prime(a_i) + \beta
\end{equation}
Where $\alpha$ and $\beta$ are constants and $c_k^\prime(a_i)$ is a rescaled version of $c_k(a_i)$. 
These constants may depend on network-specific factors, and we assume that the factor is globally applicable to all users in a given network.

With this information, the \textit{a priori} probability in Eq. \ref{equation:expected_x} can now be computed as:
\begin{eqnarray}
p(b_0, t_k) &=& \frac{\text{number of delayed reactions by } b_0 \text{ in } t_k}{\text{number of posts visible to } b_0 \text{ in } t_k}. \nonumber \\ 
            &=& \frac{r_{d,k}(b_0)}{v_k(b_0)}
\end{eqnarray}

Now we turn our attention back to the original user $a_0$. 
Let $Y_{a_0,k}$ to be the random variable representing the number of reactions that $a_0$ receives for a post created in a specific time bucket $t_k$.
We would like to find the expected number of reactions $E(Y_{a_0,k})$, which can be computed as:
\setlength{\abovedisplayskip}{2pt}
\setlength{\belowdisplayskip}{2pt}
\begin{eqnarray}\label{equation:second_deg}
E(Y_{a_0,k}) &=& E( \sum\limits_{j=0}^{m} X_{b_j, k} ) = \sum\limits_{j=0}^{m} E(X_{b_j, k} ) \nonumber \\
             &=& \sum\limits_{j=0}^{m} p(b_j, t_k) = \sum\limits_{j=0}^{m} \frac{r_{d,k}(b_j)}{(\alpha \cdot \sum\limits_{i=0}^{n} c_k^\prime(a_i) + \beta)}
\end{eqnarray}

Thus, these expected values computed from the observed $\mathcal{R}_d(u)$ and $\mathcal{C}(u)$ give us the estimates for the number of reactions received by $a_0$. 
The elements of the audience reaction profile $\mathcal{Q}(a_0)$ are hence given by:
\setlength{\abovedisplayskip}{5pt}
\setlength{\belowdisplayskip}{5pt}
\begin{equation}
q_k(a_0) = E(Y_{a_0,k}) \\
\end{equation}

Finally, we can infer the desired posting schedule $\mathcal{S}_2(a_0)$ for the user $a_0$ as the probability mass function for the discrete random variable $Y_{a_0,k}$.
Again, the elements of $\mathcal{S}_2(a_0)$ are computed from $\mathcal{Q}(a_0)$ as per Eq. \ref{equation:d_to_s}.

\subsubsection{User Weighted Schedules}
\label{sec:weighted_schedules}

In the sums computed above for the first- and second-degree schedules, all audience members are treated equally. 
However, audience members may have differing tendencies to react to the user's posts depending on their affinity to the user. 
These differences can be accounted for by associating a weight with each audience member who may react to the user, computed based on previous actions as follows:
\begin{equation}
w(a_0, b_i) = \frac{\text{total reactions received by } a_0 \text{ from } b_i }{\text{total overall reactions received by } a_0} \\
\end{equation}
Eq. \ref{equation:first_deg} can now be modified with this weight as:
\setlength{\abovedisplayskip}{0pt}
\setlength{\belowdisplayskip}{0pt}
\begin{equation}
q_k(a_0) = \sum\limits_{j=0}^{m} w(b_j, a_0) \cdot r_{d,k}(b_j)
\end{equation}
Similarly, the expected number of reactions for the second-degree schedule in Eq. \ref{equation:second_deg} can also be modified as:
\setlength{\abovedisplayskip}{0pt}
\setlength{\belowdisplayskip}{0pt}
\begin{equation}
E(Y_{a_0,k}) = \sum\limits_{j=0}^{m} w(b_j, a_0) \cdot E(X_{b_j, k} )
\end{equation}

We denote these weighted schedules as $\mathcal{S}_{1,w}(u)$ and $\mathcal{S}_{2,w}(u)$ respectively.
In Sec. \ref{section:results} we evaluate the performance of all four schedules described above.

\section{Schedule Evaluation}
\label{section:results}
In this section, we evaluate the user post schedules derived above -- $\mathcal{S}_1(u)$, $\mathcal{S}_2(u)$ and their respective weighted counterparts. We evaluate them on empirical observations of real user behavior over a $56$-day period for $0.5$ million users and more than $25$ million messages. 

\subsection{Baseline Schedules}
Because there are no previous baselines on the \textit{when-to-post} problem, we design two schedules to compare our approaches with. 
We consider all users in a given timezone and aggregate their behavior to create these baseline schedules.
Both the baselines are thus uniquely determined for each timezone and are not personalized per user.

One natural baseline can be created by observing the most frequently used time buckets for posting, aggregated across all users in each timezone $T$. 
We thus obtain our first baseline, the {\em{Most Frequently Used (MFU) Schedule}}, denoted as $\mathcal{BS}^{mfu}(T)$, with bucket values $bs^{mfu}_{i}(T)$ computed as:
\begin{equation}\label{def:most_used_bucket_schedule}
\small{
 bs^{mfu}_{i}(T) = \sum\limits_{u \in U_T} c_i(u) \big/ \sum\limits_{i=1}^{N}\sum\limits_{u \in U_T} c_i(u) \\
}
\end{equation}
where $U_T$ is the set of users in the timezone $T$.

As explained in Sec. \ref{section:derivation}, the {\em{First-Degree Reaction Schedule}} for a user is based on his first degree audience behavior.
To generate another baseline for global behavior, we simply aggregate the first-degree reaction schedules from all users in the timezone.
We call this second baseline schedule {\em{Aggregated First-Degree (AFD) Schedule}}, denoted as $\mathcal{BS}^{afd}(T)$, whose bucket values are given by:
\begin{equation}\label{def:agg_first_degree_schedule}
\small{
 bs^{afd}_{i}(T) = \sum\limits_{u \in U_T} q_{i}(u) \big/ \sum\limits_{i=1}^{N}\sum\limits_{u \in U_T} q_i(u) \\
}
\end{equation}
where $U_T$ is the set of users in the timezone $T$ who have a first-degree reaction schedule $\mathcal{Q}(u)$.

Once we have the baseline schedules, we pick the buckets with the highest values of $bs_{i}(T)$ as the best recommended times to post for users in timezone $T$.

\subsection{Evaluation metrics}
For the purposes of evaluation of schedules, we propose a {\bf{ReactionGain}} metric, which we compute as below.

Let $\mathcal{U}$ be the user sample set under consideration, observed over $M$ days.
Let us first consider a single user $u$ in this sample.
For this user $u$, we can rank the posting time buckets as recommended by a schedule $\mathcal{S}(u)$ over a period of $24$ hours, with the first bucket being the best time to post and the last one being the worst. 

For the $k^{th}$ ranked bucket as per $\mathcal{S}(u)$ we compute the average \textit{reactions per message}, $RPM(u,k)$:
\setlength{\abovedisplayskip}{0pt}
\setlength{\belowdisplayskip}{0pt}
\begin{equation}
\small{
	RPM(u, k) = \big(\sum\limits_{j=1}^{M} r_{k,j}(u)\big) \big/ \big(\sum\limits_{j=1}^{M} c_{k,j}(u)\big)
}
\end{equation}
where $r_{k,j}(u)$ and $c_{k,j}(u)$ are respectively the reactions received and the posts created by the user in the time bucket corresponding to the $k$-th rank, on the $j$-th day.
As before, we compute $r_{k,j}(u)$ as the reactions received in the first $24$ hours after the posting time.

We similarly define $RPM(u)$ as the ratio of all the reactions received to all the posts created by the user in the same $56$-day period, across all the time buckets. 
We now compute the \textit{ReactionGain}, $RG(u, k)$, for the $k$-th bucket for the user as:
\begin{equation}
RG(u, k) = \frac{RPM(u, k)}{RPM(u)}
\end{equation}
This ratio tells us the increase or decrease in reactions received by the user when she posts in a specific bucket, compared to the average reactions per message she receives.

Finally, we compute the global average reaction gain for each bucket $RG_{avg}(k)$ as the average of $RG(u, k)$ values over all the users in the sampled population $\mathcal{U}$ who created posts in that bucket.
We use this average reaction gain metric to evaluate the schedules below.

\subsection{Real-world Evaluation}
We evaluate real user behavior and measure schedule performance based on how many reactions were received when the recommended times were used.

In our experiments, we sampled $0.25$ million active users each from Twitter and Facebook from the dataset described in Sec \ref{sec:timestamp_dataset}. 
For each sampled user $u$, we compute $\mathcal{S}_1(u)$, $\mathcal{S}_2(u)$ and their corresponding weighted schedules as described in Sec. \ref{section:derivation}, for a $63$-day time period. 
We empirically chose the $\alpha$ and $\beta$ parameters to be both $1.0$, and $c_k(a_i)$ rescaled to $c_k^\prime(a_i)$ with the mean.
We then evaluate the recommended times on $25$ million messages generated by the sampled users in a $56$-day time period, with no overlap over the time period used to derive schedules. 

\begin{figure}[htbp]
  \centering
  \includegraphics[width=0.95\columnwidth]{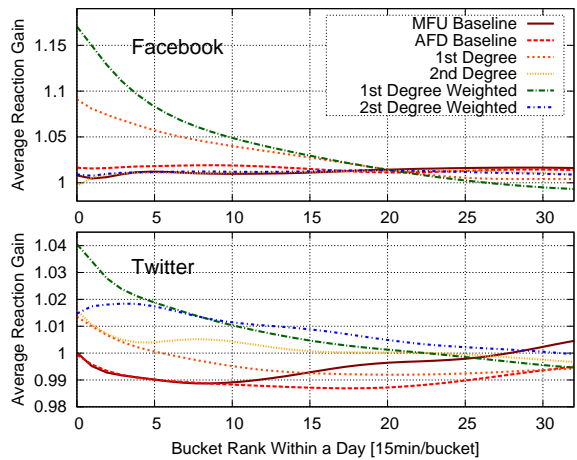}
  \caption{Average Reaction Gain for Ranked Buckets}
  \label{fig:performance_analysis}
\vspace{-0.22in}
\end{figure}

To compare the performance of the top posting times recommended by the schedules, we compute the average reaction gain $RG_{avg}(k)$ for the bucket rank $k$, for each schedule.
Fig. \ref{fig:performance_analysis} plots these values for the top $32$ buckets for a weekday\footnote{We exclude weekends here since they show diverging behavior compared to weekdays, as shown in Sec. \ref{section:behavior_analysis}, but a similar analysis can also be performed for weekends.}, for both Facebook and Twitter.


We observe from Fig. \ref{fig:performance_analysis} that the \textit{First-Degree Weighted Schedule} outperforms all the others on both Facebook and Twitter. 
On Facebook, this schedule shows a reaction gain of more than $17\%$ in the highest bucket, and on Twitter the highest gain is $4\%$. 
The second best schedule on Facebook is the \textit{First-Degree Schedule}, while that on Twitter is the \textit{Second-Degree Weighted Schedule}.
Both the \textit{MFU} and the \textit{AFD} baseline schedules show a reaction gain that is slightly above $1.0$ on Facebook, and mostly below $1.0$ on Twitter, showing that users who post according to these schedules see little to no increase in reactions received.

Both the second-degree schedules on Facebook show only a small reaction gain, very similar to the baseline schedules.
The superior performance of the first-degree schedules on Facebook suggests that second-degree effects on this network are less dominant. 
This may stem from the inherent nature of the interactions on Facebook, and the manner in which users are shown posts that they could react upon.

On Twitter, we observe that the weighted schedules for the first degree as well as second degree perform better than the baselines and the non-weighted ones. 
Thus the mutual relationships between a user and his audience members play an important role on Twitter in determining the expected reactions.
This observation highlights the importance of treating each edge in a user graph differently.


Note that a good recommended schedule should show a decreasing trend in reaction gains from the higher to the lower ranked buckets, such that posting at the higher recommended times leads to higher reaction gains.
The baseline schedules fall short in this regard, and show a decreasing trend only in the first $10$ buckets on Twitter, and none at all on Facebook. 
The global baseline schedules thus prove to be less effective in magnitude of reaction gains, as well as ordering of buckets, validating our hypothesis that personalized recommendations show better performance.

\begin{figure}[htbp]
  \centering
  \includegraphics[width=0.95\columnwidth]{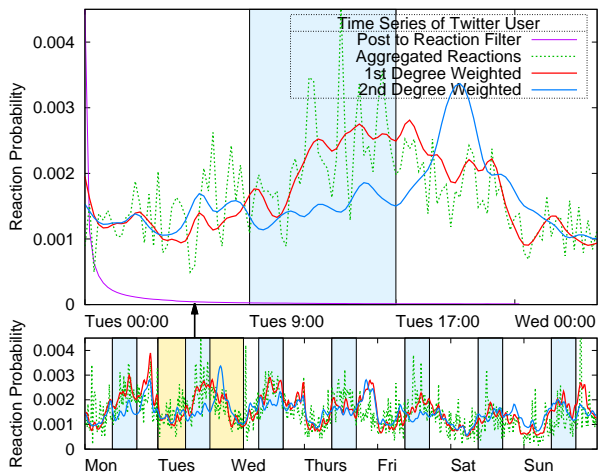}
  \caption{Example Schedules and Filter Function}
  \label{fig:time_series_example}
\vspace{-0.15in}
\end{figure}

As an example of recommended schedules, Fig. \ref{fig:time_series_example} shows the reaction profiles and schedules for a sample user on Twitter. 
The purple curve in Fig. \ref{fig:time_series_example} shows the probability distribution of post-to-reaction delay on Twitter, which is plotted by aggregating reactions observed in a $63$-day period.
Note that this function falls off steeply in the first few hours from posting time, and almost vanishes after $12$ hours.
The dashed curve plots the aggregated audience reactions for the user, without the post-to-reaction delay.
The red and the blue curves show the \textit{First-Degree Weighted Schedule} and the \textit{Second-Degree Weighted Schedule} respectively.
The recommended best times to post over one day and one week are the peaks in the plot.

\section{Conclusion and Future Work}
\label{section:conclusion}
In this study, we introduce and formulate a \textit{when-to-post} problem to find the best times to post on social networks in order to increase the number of received reactions. 

We analyze various factors that affect audience reactions on a dataset containing over a billion reactions on hundreds of millions of messages. 
We find that a majority of reactions occur within the first $2$ hours of posting times on most networks. 
Audience behavior differs significantly on different networks, with Twitter having larger reaction volumes in shorter time windows as compared to Facebook. 
We also perform location analysis and find interesting similarities and differences between cities in terms of reaction patterns.
Future studies could also study other factors such as content and topical relevance of posted messages.

Further, we present multiple approaches for deriving personalized posting schedules for users, and compare them to two baselines. 
We evaluate these schedules on empirical data from $0.5$ million real-world users and $25$ million messages observed over a $56$-day period. 
We find that the \textit{First-Degree Weighted Schedule} performs the best among all, providing a reaction gain of $17\%$ on Facebook and $4\%$ on Twitter.
Both first-degree schedules perform better on Facebook and both weighted schedules perform better on Twitter.
These schedules are deployed on a full production system that recommends posting times to millions of users daily.

We hope that this study and the accompanying dataset provided enables further research in this area.

\vspace{-0.03in}

\section{Acknowledgements}
\label{section:acknowledgements}
We thank Gaurav Ragtah, Sarah Ellinger, Tyler Singletary and Trevor D'Souza for their valuable contributions towards this study.
We also thank Sunil Rajasekar and Sateesh Chilukuri for their support throughout this work.


\bibliographystyle{abbrv}
\bibliography{igt073d-spasojevic_bibliography.bib}

\balancecolumns
\end{document}